\documentclass[aps,prc,preprint,groupedaddress,amsmath,amssymb]{revtex4}

\usepackage{graphicx}
\usepackage{dcolumn}
\usepackage{bm}

\begin{document}


\title{Effect of the tensor force in the exchange channel
on the spin-orbit splitting in $^{23}$F in the Hartree-Fock framework}


\author{Satoru Sugimoto}
\email[email: ]{satoru@ruby.scphys.kyoto-u.ac.jp} \affiliation{Kyoto
University, Kitashirakawa, Kyoto 606-8502, Japan}
\author{Hiroshi Toki}
\email[email: ]{toki@rcnp.osaka-u.ac.jp} \affiliation{Research
Center for Nuclear Physics (RCNP), Osaka University, Ibaraki, Osaka
567-0047, Japan}
\author{Kiyomi Ikeda}
\email[email: ]{k-ikeda@riken.jp}
\affiliation{The Institute of Physical and Chemical Research
(RIKEN), Wako, Saitama 351-0198, Japan}

\date{\today}

\begin{abstract}
We study the spin-orbit splitting ($ls$-splitting) for the proton $d$-orbits
in $^{23}$F in the Hartree-Fock framework with the tensor force
in the exchange channel.
$^{23}$F has one more proton around the neutron-rich nucleus $^{22}$O. A
recent experiment indicates that the $ls$-splitting for the proton $d$-orbits in $^{23}$F is
reduced
from that in $^{17}$F.  Our calculation
shows that the $ls$-splitting in $^{23}$F becomes smaller by about a
few MeV
due to the tensor force. This
effect comes from the interaction between the valence proton and the
occupied neutrons in the $0d_{5/2}$ orbit through the tensor force
and makes
the $ls$-splitting in $^{23}$F close to the experimental data.
\end{abstract}


\maketitle

\section{Introduction}
The spin-orbit splitting ($ls$-splitting) is important for the structure of
nuclei. A large $ls$-splitting between single-particle orbits with
the same orbital angular momentum is responsible for the shell
structure of nuclei \cite{mayer49}. Recently we have been obtaining much
information about unstable nuclei from various experiments. There are
experimental evidences, which indicate that the shell structure in
neutron-rich nuclei changes from that in stable nuclei. To confirm
the change of the shell structure, the information about
single-particle orbits around closed-shell or sub-closed-shell
nuclei is important. Michimasa and his collaborators studied the
proton single-particle orbits in $^{23}$F experimentally through the
proton transfer reaction \cite{michimasa07}. $^{23}$F has one more
proton
around $^{22}$O. They reported that the $ls$-splitting for the
proton $d$-orbits ($5/2^+$-$3/2^+$) is 4.06MeV, while the $ls$-splitting
for the proton $d$-orbits in $^{17}$F is 5.00MeV \cite{Tilley93,TOI1},
which is similar to the neutron $d$-orbits in $^{17}$O (5.08MeV) \cite{Tilley93,TOI1} due to the isospin symmetry.
It indicates that there is a possibility that the $ls$-splitting is
changed by the excess neutrons around $^{16}$O. The shell model
calculations reproduce the change of the $ls$-splitting from $^{17}$F
to $^{23}$F nicely
\cite{michimasa07}. In the shell model calculation the
$ls$-splitting in $^{17}$F ($^{17}$O) is an input parameter. Hence, it is
interesting to study the $ls$-splitting with a mean-field-type model,
where the $ls$-splitting is obtained self-consistently.

Hartree-Fock and Hartree-Fock-Bogoliubov calculations can now be
performed in the whole mass region over the nuclear chart. Such
mean-field
calculations can reproduce binding energies and radii of
nuclei including unstable ones using effective forces with
relatively simple forms like the Skyrme or Gogny force
\cite{vautherin72,decharge80}. In the mean field calculations the
$ls$-splitting of single-particle orbits is produced mainly by the
LS force. The $ls$-splitting of single-particle orbits and the magic
number for binding energies can be explained with the LS force
having the same strength in almost the whole mass region at least
near the stability line. Some studies show that the $ls$-splitting
in neutron-rich nuclei becomes small because the diffuseness of
the neutron density becomes large and the spin-orbit potential is
weakened \cite{dobaczewski94,lalazissis98}.

The tensor force acts on the spin of nucleon directly and should
affect the $ls$-splitting. Although the tensor force is not usually
included in the mean field calculations, some Hartree-Fock
calculations explicitly including the tensor force or the pion in
the relativistic model showed that the tensor force affects the
$ls$-splitting in spin-unsaturated nuclei
\cite{wong68,tarbutton68,bouyssy87,stancu77,Lopez00,otsuka05,otsuka06,brown06}.
Only one orbit of the spin-orbit partners is occupied in a
spin-unsaturated nucleus, while both the spin-orbit partners are
fully occupied in a spin-saturated nucleus. For example, $^{48}$Ca
is a spin-unsaturated nucleus, where the neutron $0f_{7/2}$ orbit is a
spin-unsaturated orbit and $^{40}$Ca is a spin-saturated nucleus.
Because the total spin coming from the intrinsic spin of nucleon is
zero in a spin-saturated nucleus if the wave functions of spin-orbit partners have
the same radial forms, the tensor force does not act between the
spin-saturated core and a particle or a hole around the core. In a
spin-unsaturated nucleus, the total intrinsic spin coming from the
spin-unsaturated orbit has a finite value and the tensor force
becomes active. In fact, the sizes of the $ls$-splitting for hole
orbits change from $^{40}$Ca to $^{48}$Ca and $^{16}$O to $^{22}$O
in the results of the Hartree-Fock calculations with the tensor force
or the pion \cite{tarbutton68,bouyssy87,stancu77,Lopez00,otsuka05}.
For the calcium isotopes, there is an experimental
evidence \cite{doll76} that the $ls$-splitting becomes smaller from
$^{40}$Ca to $^{48}$Ca and the order of the change is comparable to
that induced by the tensor force or the pion
\cite{bouyssy87,otsuka05}. It should be noted that
in the Hartee-Fock approximation the energy contribution from the tensor force or the pion from the direct channel
becomes zero and only that from the exchange channel has a finite value in closed-shell nuclei.

Otsuka and his collaborators discussed the effect of the tensor
force on single-particle energy in other mass regions. They nicely
reproduced the change of the splitting between $\pi 0h_{11/2}$ and
$\pi 0g_{7/2}$ in the Sb isotopes with neutron number
\cite{schiffer04} by the monopole shift induced by the tensor force
\cite{otsuka05}. They also suggested the effect of the tensor force
on the shell evolution in the neutron-rich $sd$- and $pf$-shell region
\cite{otsuka05,otsuka06}. They discussed that the neutron shell
structure changes with proton number due to the monopole interaction
between proton and neutron orbits and explained the appearance of
the magic number 16 and the disappearance of the magic number 20 in the
neutron-rich $sd$-shell region \cite{otsuka05,utsuno99,otsuka01}.
They claimed that the monopole interaction is caused by the tensor
force \cite{otsuka05,otsuka06}. To confirm such a discussion, the
direct information about a single-particle state is essential.

In this paper we perform the Hartree-Fock calculation for $^{22}$O
and $^{23}$F.
We include the tensor force and study its effect on the $ls$-splitting.
We also calculate $^{15,16,17}$O to see the effect of valence
neutrons on the $ls$-splitting and its relation to the tensor force
by comparing with $^{22}$O and $^{23}$F.
The formulation is given in
Section~\ref{sec:formulation} and the results are given in
Section~\ref{sec:result}. Section~\ref{sec:summary} is devoted to
the summary of the paper.
\section{Formulation}\label{sec:formulation}
In the preset paper we adopt two types of Hamiltonian. One includes
the 3-body force in addition to the kinetic term and the two-body
force. The other includes the density-dependent force instead of the
3-body force. The Hamiltonian with the 3-body force $H^\text{3B}$
and that with the density dependent force $H^\text{DD}$ have the
following forms,
\begin{align}
H^\text{3B}&=\sum_{i=1}^A \frac{\boldsymbol{p}_i^2}{2 M} +
\sum_{i<j}^A v(r_i,r_j) +\sum_{i<j<k}^A
v^{(3)}(r_i,r_j,r_k)-E_\text{CM}
, \\
H^\text{DD}&=\sum_{i=1}^A \frac{\boldsymbol{p}_i^2}{2 M} +
\sum_{i<j}^A v(r_i,r_j) +\sum_{i<j}^A v^\text{(DD)}(\rho;
r_i,r_j)-E_\text{CM}.
\end{align}
In the above expression, $\boldsymbol{p}$, $r$, and $M$ are the momentum,
coordinate including spin and isospin, and mass of nucleon
respectively. $A$ is a mass number. $v$ and $v^{(3)}$ are the 2-body
and 3-body potentials respectively. $v^\text{(DD)}$ is the
density-dependent potential with the one-body density $\rho$. We
subtract the energy of the center of mass motion
$E_\text{CM}=(\sum_{i}^A \boldsymbol{p}_i)^2/2 A M$.

In the Hartree-Fock calculation we assume the wave function of the
nucleus has the following form,
\begin{align}
\Psi = \mathcal{A} \prod_{\alpha} \psi_\alpha (r_\alpha)
\end{align}
with the antisymmetrization operator $\mathcal{A}$ for nucleon
coordinates. $\alpha$ labels each single-particle state and runs
over all occupied state. With the wave function the total energies
become
\begin{align}
E^\text{3B}=\sum_\alpha \langle \psi_\alpha|
\frac{\boldsymbol{p}^2}{2M}|\psi_\alpha \rangle +
\sum_{\alpha<\beta} \langle \psi_\alpha \psi_\beta | v
|\widetilde{\psi_\alpha \psi_\beta} \rangle +
\sum_{\alpha<\beta<\gamma} \langle \psi_\alpha \psi_\beta
\psi_\gamma | v^\text{(3)} |\widetilde{\psi_\alpha \psi_\beta
\psi_\gamma} \rangle
\end{align}
for $H^\text{3B}$ and
\begin{align}
E^\text{DD}=\sum_\alpha \langle \psi_\alpha|
\frac{\boldsymbol{p}^2}{2M}|\psi_\alpha \rangle +
\sum_{\alpha<\beta} \langle \psi_\alpha \psi_\beta | v
|\widetilde{\psi_\alpha \psi_\beta} \rangle + \sum_{\alpha<\beta}
\langle \psi_\alpha \psi_\beta | v^{\text(DD)}(\rho)
|\widetilde{\psi_\alpha \psi_\beta} \rangle
\end{align}
for $H^\text{DD}$, where the tildes represent the antisymmetrization.
In the above equations, $E_\text{CM}$ is dropped for simplicity. By
taking a variation of the total energy with respect to a
single-particle wave function $\psi_\alpha$, we obtain the
Hartree-Fock equation for each case:
\begin{align}
     &\frac{\boldsymbol{p}^2}{2 M} \psi_\alpha (x) + \sum_\beta \int
dy \psi_\beta^\dagger (y)
     v(x,y) \left[ \psi_\beta (y) \psi_\alpha (x) - \psi_\alpha (y)
\psi_\beta (x) \right]
     \notag \\
     &+\frac{1}{2}\sum_{\beta,\gamma} \int dy  \int dz \psi_\beta^
\dagger (y) \psi_\gamma^\dagger (z)
     v^{(3)}(x,y,z) \Bigl[
         \left\{ \psi_\beta (y) \psi_\gamma (z) - \psi_\gamma (y)
\psi_\beta (z) \right\} \psi_\alpha (x)
         \notag \\
          &+ \left\{\psi_\gamma (y) \psi_\alpha (z) - \psi_\alpha (y)
\psi_\gamma (z) \right\}\psi_\beta (x)
          + \left\{ \psi_\alpha (y) \psi_\beta (z)- \psi_\beta (y)
\psi_\alpha (z) \right\}\psi_\gamma (x)\Bigr]
=\varepsilon_\alpha \psi_\alpha (x)
\end{align}
for the three-body force case and
\begin{align}
     &\frac{\boldsymbol{p}^2}{2 M} \psi_\alpha (x) + \sum_\beta \int
dy \psi_\beta^\dagger (y)
     v(x,y) \left[ \psi_\beta (y) \psi_\alpha (x) - \psi_\alpha (y)
\psi_\beta (x) \right]
     \notag \\
     &+\sum_\beta \int dy \psi_\beta^\dagger (y)
     v^\text{(DD)}(\rho;x,y) \left[ \psi_\beta (y) \psi_\alpha (x) -
\psi_\alpha (y) \psi_\beta (x)\right]
\notag \\
&+\sum_{\beta<\gamma} \int dy \int dz  \psi_\beta^\dagger (y)
\psi_\gamma^\dagger (z)
     \frac{ \delta v^\text{(DD)}}{\delta \rho}(\rho;y,z) \frac{\delta
\rho}{ \delta \psi_\alpha^\dagger} (x) \left[ \psi_\beta (y) \psi_
\gamma (z) - \psi_\gamma (y) \psi_\beta
     (z) \right]
     =\varepsilon_\alpha \psi_\alpha (x)
\end{align}
for the density-dependent force case. In the above expression the
integrations over $y$ and $z$ include the summation over the spin and
isospin index.

In the present study we assume each single-particle state as an
eigenfunction of total spin
$\boldsymbol{j}=\boldsymbol{l}+\boldsymbol{s}$. With the assumption
a single-particle wave function can be expressed as
\begin{align}
\psi_\alpha (\boldsymbol{r}) = R_{\alpha} (r) \mathcal{Y}_{l_\alpha
j_\alpha m_\alpha} (\Omega) \zeta (\mu_\alpha),
\end{align}
where $R$ is a radial wave function, $\mathcal{Y}$ is an
eigenfunction of $\boldsymbol{j}$, and $\zeta$ is an isospin wave
function. $\alpha$ stands for node $n_\alpha$, total spin
$j_\alpha$, its projection on the $z$ axis $m_\alpha$, and isospin
$\mu_\alpha$. We do not assume the degeneracy for the orbits with the
same $n_\alpha$, $j_\alpha$, and $\mu_\alpha$ because the spherical
symmetry of a mean field is broken in odd nuclei. It means that the
states with the same $n$, $j$, and $\mu$ but different $m$'s are
allowed to have different radial wave functions.  In such a case we
need to perform an angular momentum projection to obtain a wave
function with a good angular momentum. The expectation value for the
total angular momentum $J^2$ with the wave function obtained in the
Hartree-Fock calculation for a one-particle or one-hole state does not
deviate from $j_\nu(j_\nu+1)$ largely (less than 1\%), where $j_\nu$
is the total spin of the particle or hole orbit. It indicates the
obtained wave function is almost an eigenstate of angular momentum.
Hence, we do not perform the angular momentum projection.

We approximate the density in a density-dependent force as
\begin{align}
\rho (\boldsymbol{r}) \approx \frac{1}{4\pi}\sum_\alpha
R_\alpha^\dagger (r) R_\alpha (r)
\end{align}
for calculational convenience. This expression is exact for a closed-shell nucleus with the
spherical symmetry and should be a good approximation for
a one-particle or one-hole nucleus with almost a spherical core.

We expand a radial wave function $R_\alpha(r)$ by Gaussian functions
with widths of a geometric series \cite{hiyama03}. We take 11 Gaussian
functions with the minimum width 0.5fm and the maximum width 7fm for
each single-particle state. The Hartree-Fock equation is solved by
the gradient or damped-gradient method \cite{CNP1}.
\section{Result}\label{sec:result}
In this section we apply the Hartree-Fock method to
$^{15,16,17,22}$O and $^{23}$F. We assume $^{16}$O as a closed-shell
nucleus up to the $0p$-shell and $^{22}$O as a sub-closed-shell
nucleus where the neutron $0d_{5/2}$ orbit is fully occupied in
addition to the occupied orbits in $^{16}$O. For $^{22}$O there is
the experimental evidence which suggests it has the sub-closed-shell
structure of the neutron $0d_{5/2}$ orbit \cite{thirolf00}. In the $^
{15}$O case, one neutron is
subtracted from the neutron $0p_{1/2}$ orbit or the neutron
$0p_{3/2}$ orbit in $^{16}$O. In the $^{17}$O case we add one
neutron in the $0d_{5/2}$ orbit around $^{16}$O. We do not put a
neutron in the $0d_{3/2}$ orbit in $^{17}$O because there are no
bound states in this configuration. In the $^{23}$F case we add a proton
in the $0d_{5/2}$ or $0d_{3/2}$ orbit around $^{22}$O.

As for the effective interaction, we adopt the modified Volkov force
No.~1 (MV1)\cite{ando80} for the central part and the G3RS force
\cite{tamagaki68} for the tensor part. We also include the Coulomb
force. The G3RS force is determined
to reproduce the nucleon-nucleon scattering data and, therefore, the
tensor
force in the G3RS force is the one in the free space. For the strength of the
tensor force in the nuclear medium we do not have a definite
guideline at present. The
effective interaction obtained from the $G$-matrix theory has a
tensor part with a strength comparable to the tensor force in the free
space \cite{sprung72,kohno75,myo07,otsuka05} at least in the region
where the relative distance is greater than about 0.8fm. We use the
tensor force in the free space in the present calculation but we need a further investigation to determine the
strength of the tensor force to be used in a mean field calculation.
It should be noted that the difference in the short range ($r<0.8$fm)
does not influence the tensor force matrix elements significantly
\cite{myo07}. As
for the LS force we take the $\delta$-type LS force
\cite{vautherin72,decharge80}:
\begin{align}
i W_0 (\boldsymbol{\sigma}_1+\boldsymbol{\sigma}_2)\cdot
\overleftarrow{\boldsymbol{k}} \times
\delta(\boldsymbol{r}_1-\boldsymbol{r}_2)
\overrightarrow{\boldsymbol{k}}.
\end{align}
The Majorana parameter in the MV1 force is fixed to 0.59, which
is determined to reproduce the binding energy of $^{16}$O. $W_0$ in
the LS force is taken as 115MeVfm$^{5}$, which is the same as in the
Gogny D1 force and is determined to reproduce the $ls$-splitting
for the $0p$ orbits in $^{15}$O \cite{decharge80}.

\begin{table}[htb]
\caption{\label{tbl:16O} Total energy ($E_\text{TOT}$), kinetic
energy ($T$) and potential energy ($V$) of $^{16}$O, $^{17}$O, and
$^{15}$O. $V_\text{LS}$ and $V_\text{T}$ are the contributions from
the LS and tensor forces to the potential energy. Those are give
in MeV. $R_\text{c}$ and $R_\text{m}$ are the charge and matter
radii in fm. The last row shows the differences of energies between
$^{15}$O ($0p_{3/2}^{-1}$) and $^{15}$O ($0p_{1/2}^{-1}$). In the
parentheses the experimental data are given.}
\begin{ruledtabular}
\begin{tabular}{crrrrrrrrrr}

            &$E_\text{TOT}$ &    &      $T$ &          $V$ &
$V_\text{LS}$ &         $V_\text{T}$ &         $R_\text{c}$  &
&         $R_\text{m}$& \\
\hline
        $^{16}$O &    $-$128.3 &($-$127.6\footnote[1]{Reference \cite
{audi03}.}) &     233.8  &    $-$362.0  &      $-$1.0  &       0.0  &
        2.71&
        (2.730(25)\footnote[2]{Reference \cite{devries87}.})
&      2.58 &(2.54(02)\footnote[3]{Reference \cite{ozawa01}.})
   \\

$^{17}$O ($0d_{5/2}$) &    $-$132.3 & ($-$131.8\footnotemark[1])  &
254.7 & $-$387.0&
$-$4.1  &       0.0  &      2.72 & (2.662(26)\footnotemark[2]) &
2.64 & (2.59(05)\footnotemark[3]) \\

$^{15}$O ($0p_{1/2}^{-1}$) &    $-$110.2 & ($-$112.0\footnotemark[1])  &
219.6 & $-$329.7 &      $-$4.9  &      $-$0.1  &      2.70&   &      2.55&
(2.44(04)\footnotemark[3])
\\

$^{15}$O ($0p_{3/2}^{-1}$) &    $-$104.5 & &     212.4  &    $-$316.9
&       0.9  &       0.0  &      2.74&  &      2.59&  \\

       $\Delta(0p_{3/2}^{-1}-0p_{1/2}^{-1})$ &      5.7  & (6.18
\footnote[4]{Reference \cite{Ajzenberg91,TOI1}.}) &       $-$7.2  &
12.8  &      5.8  &      0.1  &            &            \\

\end{tabular}
\end{ruledtabular}
\end{table}
In Table~\ref{tbl:16O}, the results for $^{16}$O, $^{17}$O, and
$^{15}$O are summarized. The experimental data are also given in the
parentheses if available. The potential energy from the tensor force
becomes quite small because $^{16}$O is a LS-closed-shell nucleus.
In the LS-closed-shell nucleus both the spin-orbit partners are
completely occupied. Hence, the LS-closed-shell nucleus is a spin-saturated nucleus.
The LS-closed-shell nucleus does not have
a finite total orbital angular momentum
and a finite total spin angular momentum. The tensor force
consists of the rank 2 tensors of the orbital and spin angular
momenta. Thus, the tensor force does not work between the
LS-closed-shell nucleus and a particle or a hole around it, because a
particle or hole has a spin angular momentum 1/2. In the last row
the energy differences between $^{15}$O ($0p_{3/2}^{-1}$) and
$^{15}$O ($0p_{1/2}^{-1}$) are shown. It corresponds to
the $ls$-splitting for the $0p$ orbits. It is about 10\% smaller than
the experimental value. The contribution from the LS force is 5.8MeV
and is almost the same as the total $ls$-splitting. It indicates
that the $ls$-splitting is mainly produced by the LS force. The
large contribution from the kinetic energy is almost canceled out with
the contributions from the central and three-body forces. In
$^{15}$O the effect of the tensor force on the $ls$-splitting is
negligible.

\begin{table}[htb]
\caption{\label{tbl:22O}Total energy ($E_\text{TOT}$), kinetic
energy ($T$) and potential energy ($V$) of $^{22}$O and $^{23}$F.
$V_\text{LS}$ and $V_\text{T}$ are the contributions from
the LS and tensor forces to the potential energy. Those are give
in MeV. $R_\text{c}$ and $R_\text{m}$ are the charge and matter
radii in fm. The last row is the differences of energies between
$^{23}$F ($0d_{3/2}$) and $^{23}$F ($0d_{5/2}$). In the parentheses
the experimental data are given.}
\begin{ruledtabular}
\begin{tabular}{crrrrrrrrrr}
&$E_\text{TOT}$& &$T$ & $V$ & $V_\text{LS}$ & $V_\text{T}$ & $R_\text
{c}$ & $R_\text{m}$ & \\
\hline
        $^{22}$O &    $-$161.8 &($-$162.0\footnote[1]{Reference \cite
{audi03}.})
&     361.4  &    $-$523.2  &     $-$20.8  &       1.9  &      2.74  &
2.85 & (2.88(06)\footnote[2]{Reference \cite{ozawa01}.})
\\

$^{23}$F ($0d_{3/2}$) &    $-$166.5 & &     376.4  &    $-$542.8  &
$-$16.3  &       0.1  &      2.89  &      2.90 & \\

$^{23}$F ($0d_{5/2}$) &    $-$170.7 &($-$175.3\footnotemark[1]) & 383.9
& $-$554.5 & $-$24.1  &       3.2  &      2.84        &2.87
&(2.79(04)\footnotemark[2])
\\

       $\Delta (0d_{3/2}-0d_{5/2}$) &       4.2 & (4.06\footnote[3]
{Reference \cite{michimasa07}.}) &      $-$7.5  &      11.7  &
7.8  &      $-$3.1  &            &  &          \\

\end{tabular}
\end{ruledtabular}
\end{table}
In Table~\ref{tbl:22O} the results for $^{22}$O and $^{23}$F are
summarized. Although the binding energy of $^{23}$F ($0d_{5/2}$)
(the ground state) is about 5MeV smaller than the experimental value,
it probably does not affect our discussion on the $ls$-splitting. In
$^{22}$O the neutron $0d_{5/2}$ orbit around the
$^{16}$O core is fully occupied. Because the spin-orbit partner, the
neutron $0d_{3/2}$ orbit, is empty, $^{22}$O is a spin-unsaturated
nucleus. Hence, $^{22}$O has a finite total orbital angular momentum
and a finite total spin angular momentum, and the expectation value for the
tensor potential energy in $^{23}$F becomes finite. In $^{22}$O, the energy
contributions from the LS force and the tensor force are $-20.8$MeV and
1.9MeV respectively. In $^{23}$F a proton is added to $^{22}$O. If
the proton is put in the $0d_{3/2}$ orbit the absolute value of the
LS potential energy
becomes small by 4.5MeV and if the proton is put in the $0d_{5/2}$
orbit that of the LS potential energy becomes large by 3.3MeV. In
contrast,
the tensor potential energy becomes small by 1.8MeV when the proton
is in the $0d_{3/2}$ orbit and becomes large by 1.3MeV when the
proton is in the $0d_{5/2}$ orbit. As a result, the contribution to
the $ls$-splitting for the proton $0d$ orbits in $^{23}$F from the
LS force is 7.8MeV and that from the tensor force is $-3.1$MeV. The
sum of them is 4.5MeV. The relatively small $ls$-splitting 4.2MeV
after adding the contributions from the kinetic and other potential energies,
which is close to the experimental value, is realized by the
cancelation between the contributions from the LS force and the tensor force.

\begin{table}[htb]
\caption{\label{tbl:dLSdT} Differences of the LS potential energy
($\Delta(V_\text{LS})$) and the tensor potential energy
($\Delta(V_\text{T})$) between one-particle nuclei and their core
nuclei. Those are given in MeV.}
\begin{ruledtabular}
\begin{tabular}{crr}

            &      $\Delta(V_\text{LS})$ & $\Delta (V_\text{T})$ \\
\hline
$^{17}\text{O} (0d_{5/2})-^{16}\text{O}$ &      $-$3.0  &       0.0  \\

$^{23}\text{F} (0d_{3/2})-^{22}\text{O}$ &       4.5  &      $-$1.8  \\

$^{23}\text{F} (0d_{5/2})-^{22}\text{O}$ &      $-$3.3  &       1.3  \\

\end{tabular}
\end{ruledtabular}
\end{table}
The energy differences between one-particle states and their
corresponding
cores are shown in Table~\ref{tbl:dLSdT}. The LS potential energies
from the cores for the $0d_{5/2}$ orbit are $-3.0$MeV in $^{17}$O and
$-3.3$MeV in $^{23}$F.
The LS potential energy for the $0d_{3/2}$ orbit in $^{23}$F is
smaller as expected from that for the $0d_{5/2}$ orbit
($3.3 \times (2+1)/2 \approx 5.0$MeV). It is probably due to a weak
binding of the
$0d_{3/2}$ orbit compared to the $0d_{5/2}$ one. The contribution
from the tensor force to the splitting for the $0d$ orbits in $^{23}$F
is about a half
of that from the LS force with the opposite sign as discussed in the
previous section. The results for $^{17}$O and $^{23}$F in
Table~\ref{tbl:dLSdT} indicate that the contribution to the
$ls$-splitting from the LS force mainly comes from the $^{16}$O core
and that from the tensor force comes from the excess neutron orbit
(the neutron $0d_{5/2}$ orbit).

\begin{table}[htb]
\caption{\label{tbl:VTEO} Potential energy contributions from the
triplet-even tensor force ($V_\text{T}^{3\text{E}}$) and the
triplet-odd tensor force ($V_\text{T}^{3\text{O}}$) in MeV. In the
last two rows, the differences between $^{23}$F ($0d_{3/2}$ or
$0d_{5/2}$) and $^{22}$O are given.}
\begin{ruledtabular}
\begin{tabular}{crr}

       &  $V_\text{T}^{3\text{E}}$ &   $V_\text{T}^{3\text{O}}$ \\
\hline
   $^{22}$O &  0.1  &  1.8  \\

$^{23}$F (0$d_{3/2}$) & $-$1.3  &  1.4  \\

$^{23}$F (0$d_{5/2}$) &  1.0  &  2.1  \\

$\Delta (^{23}\text{F} (0d_{3/2})-^{22}\text{O})$ & $-$1.4  & $-$0.4  \\

$\Delta (^{23}\text{F} (0d_{5/2})-^{22}\text{O})$ &  1.0  &  0.3  \\

\end{tabular}
\end{ruledtabular}
\end{table}
In Table~\ref{tbl:VTEO}, the contributions to the tensor potential
energy from the triplet-even and triplet-odd parts are shown
separately. In
$^{22}$O the tensor potential energy mainly comes from the
triplet-odd part. It is natural because only the neutron $0d_{5/2}$
orbit is occupied and there are no valence protons around the
$^{16}$O core. In $^{23}$F the contribution from the triplet-even
part is comparable to that from the triplet-odd part for the
$0d_{3/2}$ orbit and they have the opposite sign.
For the $0d_{5/2}$ orbit the contribution from the triplet-even
part is smaller than that from the triplet-odd
part and they have the same sign. To see the effect of the tensor force
on the valence proton, the energy differences between $^{23}$F and
$^{22}$O are shown in the table. The differences are dominated by
the triplet-even part. It means that the contribution to the
$ls$-splitting from the tensor force mainly comes from the
triplet-even tensor force.

\begin{table}[htb]
\caption{\label{tbl:deltad} $ls$-splitting for the proton $d$-orbits in $^{23}$F with various effective interactions (see the
text). $\Delta (V_\text{LS})$, $\Delta (V_\text{T})$, and $\Delta
(\text{others})$ are the contributions to the $ls$-splitting from
the LS force, the tensor force, and the other forces including the
kinetic term respectively. Those are given in MeV. The experimental
value for $\Delta(0d_{3/2}-0d_{5/2})$ = 4.06MeV.}

\begin{ruledtabular}
\begin{tabular}{crrrr}

            &      $\Delta (0d_{3/2}-0d_{5/2})$  &        $\Delta (V_
\text{LS})$ &         $\Delta (V_\text{T}$) &  $\Delta (\text{others})
$ \\
\hline
MV1 &       4.2  &       7.8  &      $-$3.1  &      $-$0.5  \\
MV1 without $V_\text{T}$ &       7.2  &       8.3  &       0.0
&      $-$1.1  \\
      Gogny D1S &       8.5  &       9.4  &       0.0  &      $-$0.9  \\
        M3Y-P2 &       7.6  &       9.2  &      $-$0.4  &      $-$1.2  \\
GT2  &       8.2  &      12.2  &      $-$3.3  &      $-$0.7  \\
\end{tabular}
\end{ruledtabular}
\end{table}
Finally we compare the $ls$-splitting calculated with other
effective interactions with our result discussed above (MV1) in
Table~\ref{tbl:deltad}. We also show the result without the tensor
force (MV1 without $V_\text{T}$). The Gogny D1S
force \cite{berger91} does not have a tensor part and a stronger LS
part ($W_0$=130MeVfm$^5$) than one we adopted above. The M3Y-P2
force \cite{nakada03} has a weak tensor part and an LS part
comparable to the Gogny D1S force. The GT2 \cite{otsuka06,matsuo03}
force has a tensor part comparable to that in the free space and a
strong LS part ($W_0$=160MeVfm$^5$). While the rather schematic
form of the tensor force is adopted in Ref.~\onlinecite{otsuka06},
we replace the tensor part of the GT2 force with the G3RS force we
used above. The sizes
of the $ls$-splitting for the MV1 force without the tensor force,
the Gogny force, and the M3Y-P2 force are large compared to the
experimental value.  It indicates that the relatively strong tensor
force comparable to that in the free space is needed to reproduce the
$ls$-splitting in $^{23}$F. Although the GT2 force has a strong
tensor part, it gives quite large splitting. It is due to the
strong LS part of the GT2 force. The contribution from the LS force
to the $ls$-splitting is much larger than those with other effective
forces. In fact, the $ls$-splitting for the $0p$ orbits in $^{15}$O
with the GT2 force is 8.3MeV. It is much larger than the
experimental value.  It indicates that the proper strength of the LS
force, which give the reasonable $ls$-splitting in $^{15}$O is
needed to reproduce the $ls$-splitting in $^{23}$F.

The tensor force also induces a 2-particle--2-hole ($2p2h$)
correlation, which cannot be treated in a usual mean field
calculation. The $2p2h$ correlation by the tensor force produces the
large attractive energy in nuclei \cite{bethe71,akaishi86}. Recently
we developed a mean field framework which can treat the $2p2h$
tensor correlation by introducing single-particle states with charge
and parity mixing \cite{toki02,ogawa04,sugimoto04,sugimoto07}. We
applied the extended mean field model to sub-closed-shell oxygen
isotopes \cite{sugimoto07} and found that the potential energy from
the tensor force is comparable to that from the LS force. The
importance of the $2p2h$ tensor correlation for the $ls$-splitting
is indicated in other studies \cite{terasawa60,ando81,myo05,ogawa04}.
It is interesting to study the effect of the $2p2h$ tensor
correlation on the $ls$-splitting with our extended mean filed model. Because our
calculation showed that the excess neutrons around $^{16}$O do not
contribute to the $2p2h$ tensor correlation strongly \cite{sugimoto07},
the Hartree-Fock calculation seems to be sufficient as the first step.
\section{Summary}\label{sec:summary}
We have performed the Hartree-Fock calculation with the tensor force for
$^{15}$O, $^{16}$O, $^{17}$O, $^{22}$O, and $^{23}$F to study the
effect of the tensor force on the $ls$-splitting.

The tensor force does not affect the $ls$-splitting for the $0p$
orbits in $^{15}$O because $^{16}$O is a LS-closed-shell nucleus.
The $ls$-splitting is almost produced by the LS force in $^{15}$O.

In $^{22}$O, the neutron $0d_{5/2}$ orbit is fully occupied. It
gives the finite expectation value for the tensor force in $^{22}$O.
In $^{23}$F a proton is added to $^{22}$O. The LS force works to
provide the $ls$-splitting for the proton $0d$-orbits in $^{23}$F by
7.8MeV. In contrast, the tensor force reduces the $ls$-splitting by
3.1MeV. The effect of the tensor force mainly comes from the
occupied neutron $0d_{5/2}$ orbit. The resulting $ls$-splitting of
4.2MeV
close to the experimental data is realized by the cancelation
between the effects of the LS force and the tensor force. The
contribution from the tensor force to the $ls$-splitting in $^{23}$F mainly comes
from the triplet-even part of the tensor force.

We have compared the results with various effective interactions with
and without
the tensor force. The effective interaction without the tensor force
or with the weak tensor force does not explain the experimental
value for the $ls$-splitting for the proton $0d$-orbits in $^{23}$F.
Our study indicates that the LS and tensor forces with reasonable
strengths are needed
to reproduce the $ls$-splitting in $^{15}$O and $^{23}$F,
simultaneously.
\begin{acknowledgments}
We acknowledge fruitful discussions with Prof. H.~Horiuchi on the
role of the tensor force in light nuclei. This work is supported by
the Grant-in-Aid for the 21st Century COE ``Center for Diversity and
Universality in Physics'' from the Ministry of Education, Culture,
Sports, Science and Technology (MEXT) of Japan. A part of the
calculation of the present study was performed on the RCNP computer
system.
\end{acknowledgments}


\begin{thebibliography}{99}
\bibitem{mayer49}
M.~G.~Mayer, Phys. Rev. \textbf{75}, 1969 (1949);
O.~Haxel, J.~H.~D.~Jensen, and H.~E.~Suess, Phys. Rev. \textbf{75}, 1766 (1949).
\bibitem{michimasa07}
S.~Michimasa \textit{et al.}, Nucl. Phys. A (to be published).
\bibitem{TOI1}
R.~B.~Firestone, \textit{Table of Isotopes} (John Wiley \& Sons, New
York, 1996), Vol.~1.
\bibitem{Tilley93} D.~R.~Tilley, H.~R.~Weller and C.~M.~Cheves, Nucl.
Phys. A \textbf{565}, 1 (1993).
%
\bibitem{vautherin72} D.~Vautherin and D.M.~Brink, Phys. Rev. C
\textbf{5}, 626 (1972).
\bibitem{decharge80} J.~Decharg\'e and D.~Gogny, Phys. Rev. C
\textbf{21}, 1568 (1980).
%
\bibitem{dobaczewski94} J.~Dobaczewski, I.~Hamamoto, W.~Nazarewicz,
and J.~A.~Sheikh, Phys. Rev. Lett. \textbf{72}, 981 (1994).
\bibitem{lalazissis98} G.~A.~Lalazissis, D.~Vretenar, W.~P\"oschl, and
P.~Ring, Phys. Lett. B \textbf{418}, 7 (1998).
\bibitem{wong68}
C.~W.~Wong, Nucl. Phys. A \textbf{108}, 481 (1968).
\bibitem{tarbutton68}
R.~M.~Tarbutton and K.~T.~R.~Davies, Nucl Phys. A \textbf{120},1
(1968).
%
\bibitem{stancu77} Fl.~Stancu, D.~M.~Brink, and H.~Flocard, Phys.
Lett. B \textbf{68}, 108 (1977).
\bibitem{bouyssy87}
A.~Bouyssy, J.-F.~Mathiot, N.~Van~Giai, and S.~Marcos, Phys. Rev. C
\textbf{36}, 380 (1987).
%
\bibitem{Lopez00} M.~L\"opez-Quelle, N.~Van~Giai, S.~Marcos, and
L.~N.~Savushkin, Phys. Rev. C \textbf{61}, 064321 (2000).
%
\bibitem{otsuka05} T.~Otsuka, T.~Suzuki, R.~Fujimoto, H.~Grawe, and
Y.~Akaishi, Phys. Rev. Lett. \textbf{95}, 232502 (2005).
\bibitem{otsuka06} T.~Otsuka, T.~Matsuo, and D.~Abe, Phys. Rev.
Lett. \textbf{97}, 162501 (2006).
\bibitem{brown06} B.~A.~Brown, T.~Duguet, T.~Otsuka, D.~Abe, and
T.~Suzuki, Phys. Rev. C. \textbf{74}, 061303(R) (2006).
%
\bibitem{doll76} P.~Doll, G.~J.~Wagner, K.~T.~Kn\"opfle, and
G.~Mairle, Nucl. Phys. A \textbf{263}, 210 (1976).
%
\bibitem{schiffer04} J.~P.~Schiffer \textit{et al.}, Phys. Rev. Lett.
\textbf{92}, 162501 (2004).
%
\bibitem{utsuno99} Y.~Utsuno, T.~Otsuka, T.~Mizusaki, and M.~Honma,
Phys. Rev. C \textbf{60},  054315 (1999).
%
\bibitem{otsuka01} T.~Otsuka, R.~Fujimoto, Y.~Utsuno, B.~A.~Brown,
M.~Honma, and T.~Mizusaki, Phys. Rev. Lett. \textbf{87}, 082502
(2001).
%
%
%
%
%
\bibitem{hiyama03}
E.~Hiyama, Y.~Kino, and M.~Kamimura, {Prog. Part. Nucl. Phys.}
\textbf{51}, 223 (2003).
%
\bibitem{CNP1}
P.-G.~Reinhard, in \textit{Cmpulationl Nuclear Phyiscs 1} (edited by
K.~Langanke, J.~A.~Maruhn, and S.~E.~Koonin, Springer-Verlag,
Berlin, 1991) Chapter 2.
%
\bibitem{thirolf00} P.~G.~Thirolf \textit{et al.}, Phys. Lett. B
\textbf{485}, 16 (2000).
%
\bibitem{ando80} T.~Ando, K.~Ikeda, and A.~Tohsaki-Suzuki, Prog.
Theor. Phys. \textbf{64}, 1608 (1980).
%
%
\bibitem{tamagaki68}
R.~Tamagaki, Prog. Theor. Phys. \textbf{39}, 91 (1968).
%
\bibitem{sprung72} D.~W.~Sprung, Nucl. Phys. A \textbf{182}, 97
(1972).
%
\bibitem{kohno75}
M.~Kohno, S.~Nagata, and N.~Yamaguchi, Prog. Theor. Phys. Suppl.
\textbf{65}, 200 (1975).
%
%
\bibitem{myo07}
T.~Myo, S.~Sugimoto, K.~Kat\={o}, H.~Toki, and K.~Ikeda, Prog.
Theor. Phys.  \textbf{117}, 257 (2007).
%
\bibitem{audi03}
G.~Audi, A. H.~Wapstra and C.~Thibault, Nucl. Phys. A \textbf{729},
337 (2003).
\bibitem{devries87}
H.~de~Vries, C.~W.~de~Jager, and C.~de~Vries, At. Data Nucl. Data
Tables \textbf{36}, 495 (1987).
\bibitem{ozawa01}
A.~Ozawa, T.~Suzuki and I.~Tanihata, Nucl. Phys. A \textbf{693}, 32
(2001).
\bibitem{Ajzenberg91} F.~Ajzenberg-Selove, Nucl. Phys. A \textbf
{523}, 1 (1991).
%
%
%
%
\bibitem{berger91} J.~F.~Berger, M.~Girod, and D.~Gogny, Comput.
Phys. Commun. \textbf{63}, 365 (1991).
%
\bibitem{nakada03} H.~Nakada, Phys. Rev. C \textbf{68}, 014316
(2003).
%
\bibitem{matsuo03} T.~Matsuo, Ph.~D. thesis, the University of
Tokyo, 2003.
%
\bibitem{bethe71}
H.~A.~Bethe, Annu. Rev. Nucl. Sci. \textbf{21}, 93 (1971).
%
\bibitem{akaishi86}
Y.~Akaishi, in {Cluster Models and Other Topics}, edited by
T.~T.~S.~Kuo and E.~Osnes (World Scientific, Singapore, 1986),
p.~259.
%
\bibitem{toki02} H.~Toki, S.~Sugimoto, and K.~Ikeda,
Prog. Theor. Phys. \textbf{108} 903 (2002).
%
\bibitem{sugimoto04} S.~Sugimoto, K.~Ikeda, and H.~Toki, Nucl. Phys.
A \textbf{740}, 77 (2004).
%
\bibitem{ogawa04} Y.~Ogawa, H.~Toki, S.~Tamenaga, H.~Shen,
A.~Hosaka, S.~Sugimoto, and K.~Ikeda, Prog. Theor. Phys.
\textbf{111}, 75 (2004).
%
%
\bibitem{sugimoto07} S.~Sugimoto, K.~Ikeda, and H.~Toki, Phys. Rev.
C \textbf{75}, 014317 (2007).
%
\bibitem{terasawa60}
S.~Takagi, W.~Watari, and M.~Yasuno, Prog. Theor. Phys. \textbf{22},
549 (1959); T.~Terasawa, Prog. Theor. Phys. \textbf{23}, 87 (1960);
A.~Arima and T.~Terasawa, Prog. Theor. Phys. \textbf{23}, 115
(1960).
%
\bibitem{ando81}
K.~And\=o and H.~Band\=o, Prog. Theor. Phys. \textbf{66}, 227
(1980).
%
\bibitem{myo05} T.~Myo, K.~Kat\={o}, and K.~Ikeda, Prog. Theor.
Phys. \textbf{113}, 763 (2005).
%
\end{thebibliography}

\end{document}